\documentclass[superscriptaddress,showpacs,preprintnumbers,amsmath,amssymb,twocolumn]{revtex4}

\usepackage{epsfig}
\usepackage{dcolumn}
\usepackage{bm}
\usepackage{latexsym}

\begin{document}
\preprint{APS/123-QED}

\title{Optical orientation of Mn$^{2+}$ ions in GaAs}

\author{I.~A. Akimov}
 \affiliation{Experimentelle Physik 2, Technische Universit\"at Dortmund, 44221 Dortmund, Germany}
 \affiliation{A.F. Ioffe Physical-Technical Institute, Russian Academy of Sciences, 194021 St. Petersburg, Russia}
\author{R.~I. Dzhioev}
\author{V.~L. Korenev}
\author{Yu.~G. Kusrayev}
\author{V.~F. Sapega}
 \affiliation{A.F. Ioffe Physical-Technical Institute, Russian Academy of Sciences, 194021 St. Petersburg, Russia}
\author{D.~R. Yakovlev}
 \affiliation{Experimentelle Physik 2, Technische Universit\"at Dortmund, 44221 Dortmund, Germany}
 \affiliation{A.F. Ioffe Physical-Technical Institute, Russian Academy of Sciences, 194021 St. Petersburg, Russia}
\author{M. Bayer}
 \affiliation{Experimentelle Physik 2, Technische Universit\"at Dortmund, 44221 Dortmund, Germany}
\date{\today}

\begin{abstract}
We report on optical orientation of Mn$^{2+}$ ions in bulk GaAs under application of weak longitudinal magnetic fields ($B \leq 100$~mT). A manganese spin
polarization of 25\% is directly evaluated using spin-flip Raman scattering. The dynamical Mn$^{2+}$ polarization occurs due to the s-d exchange interaction with
optically oriented conduction band electrons. Time-resolved photoluminescence reveals a nontrivial electron spin dynamics, where the oriented Mn$^{2+}$ ions tend to
stabilize the electron spin.
\end{abstract}



\maketitle

Electron spin interactions in semiconductors are in the focus of current attention as they are of interest for various applications in electronics and quantum
information. A pathway to spin manipulation is to implement magnetic ions in semiconductors and exploit the strong exchange interaction between an electron (hole)
and a magnetic ion (for example Mn). A spectacular example to that end is a quantum dot (QD) with a single Mn ion \cite{Besombes04, Krebs07}. Another model
realization has been achieved for GaAs doped with Mn acceptors of low concentration, where a photoexcited electron is localized on a residual donor in vicinity of a
Mn acceptor \cite{Astakhov08, Akimov10}. Optical orientation of manganese acceptors and their spin control are attractive for investigation of various nontrivial
scenarios of electron spin dynamics in semiconductors, e.g. multi-exponential spin decay or spin precession in effective exchange fields. Optical orientation of a
single Mn$^{2+}$ ion in a II-VI QD has been recently reported \cite{Besombes09, Goryca09}. However, the question whether this is feasible in III-V materials, such
as GaAs, has remained open \cite{Akimov10}.

Here we report on the direct observation of optical spin orientation of Mn$^{2+}$ ions in bulk GaAs under optical illumination with circular polarized light in
longitudinal magnetic fields (Faraday geometry). In spin-flip Raman scattering (SFRS) we observe spectral lines corresponding to spin-flip processes of the ionized
Mn acceptor. The asymmetry of Stokes and anti-Stokes (SAS) line intensities under circular polarized excitation opens a way for direct evaluation of the manganese
spin polarization, which can reach 25\%. Time-resolved photoluminescence (TRPL) reveals a nontrivial electron spin dynamics resulting from the stabilizing feedback
of the oriented manganese on the electron spin. The results of SFRS and TRPL agree well, supporting clearly optical manganese orientation in weak magnetic fields.

We investigate bulk GaAs doped with Mn acceptors with a
concentration of 8$\times10^{17}$~cm$^{-3}$ \cite{Astakhov08,
Akimov10}. The acceptors are partially compensated by residual
shallow donors with a concentration of about $N_D \sim
10^{16}$~cm$^{-3}$ \cite{Footnote2}. This results in ionization of
the acceptors located in the vicinity of donors if not illuminated
by light. Detailed descriptions of the experimental techniques can
be found in Ref.~\cite{Sapega01} for SFRS and in
Ref.~\cite{Akimov10} for TRPL.

In order to detect manganese polarization we exploit SFRS technique, which selectively monitors the changes of Mn$^{2+}$ spin projection by $n=1,2..2I$ with Mn spin
$I=5/2$. It enables to measure both the Zeeman splitting $n \mu_B g B$ and the amplitudes of SAS components. Here $\mu_B$ is the Bohr magneton, and $g=2$ is the Mn
- Lande factor. The ratio between the Stokes and anti-Stokes line intensities, $\mathcal{I}^S$ and $\mathcal{I}^{AS}$, respectively is given not only by the
selection rules but also by the Mn$^{2+}$ spin polarization $P_M$. It can be shown that in Faraday geometry for circular co-polarized excitation and detection with
helicity $\sigma$ and for a small manganese polarization ($P_M \ll 1$) the SAS asymmetry parameter of SFRS,
\begin{equation}
\label{eq:SFRSasymmetry}
\eta_n^\sigma=\frac{\mathcal{I}^{S}_{n\sigma}-\mathcal{I}^{AS}_{n\sigma}}{\mathcal{I}^{S}_{n\sigma}+\mathcal{I}^{AS}_{n\sigma}}=-\frac{3n}{2(I+1)}P_M(\sigma,B),
\end{equation}
is independent on the specific Mn spin-flip mechanism. In this case $\eta_n^\sigma$ directly monitors the spin polarization of manganese. In
Eq.~(\ref{eq:SFRSasymmetry}) we assumed: (i) the spectral width of the spin-flip resonance profile is larger than the Zeeman spin splitting; (ii) Raman scattering
without conservation of the total angular momentum of manganese and the exciton in the intermediate state is possible. Such transitions are allowed in presence of
anisotropic exchange interaction between the exciton and the manganese $3d^5$ electrons, i.e. if the symmetry of the system is reduced \cite{Ivchenko92,Sapega94}.

The manganese polarization $P_M(B,P_e)$ depends not only on the external magnetic field due to the thermal population of the Zeeman sublevels, but also on the
non-equilibrium electron polarization $P_e$ due to the dynamical polarization of Mn spins by electrons (holes are not oriented in bulk GaAs \cite{OO}). This process
is similar to dynamical polarization of the nuclei in semiconductors (Overhauser effect) and has been considered for II-VI diluted magnetic semiconductors
\cite{Dietl89}. The dynamical polarization changes sign if the helicity of circular polarized excitation, $\sigma^+$ or $\sigma^-$, is reversed. Therefore, it is
necessary to measure the two asymmetry parameters $\eta_n^{\sigma^+}$ and $\eta_n^{\sigma^-}$ for the two polarizations.

\begin{figure}
 \begin{minipage}{8.2cm}
  \epsfxsize=7.5 cm
  \centerline{\epsffile{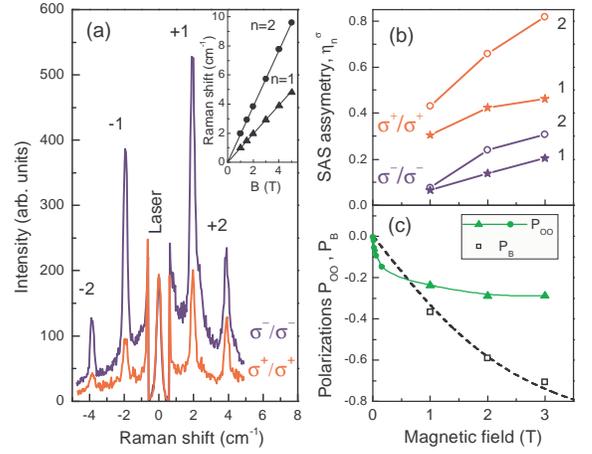}}
  \caption{\label{fig:SFRSpectra}
(a) SFRS spectra at $B=2$~T for a bath temperature 2~K. The numbers next to the lines indicate the $n$ with positive sign ($+$) for Stokes and negative ($-$) for
anti-Stokes components. Inset shows the fan charts of the lines. (b) Magnetic field dependence of SAS asymmetry $\eta_n^\sigma$ after Eq.~(\ref{eq:SFRSasymmetry}).
Solid stars (open circles) correspond to $n=1$ ($n=2$). (c) Polarizations $P_{OO}$ (solid triangles) and $P_B$ (open squares) extracted from $\eta_n^\sigma$ and
averaged over $n$. Solid circles give Mn polarization deduced from TRPL [see also Fig.~\ref{fig:TPRLplateau}(c)]. $P_{OO}$ is given for $\sigma^+$ excitation. Solid
line is a guide to the eye. Dashed line is a fit of $P_B$ with Brillouin function for $I=5/2$ and $T=4.4$~K. }
  \end{minipage}
\end{figure}

Figure \ref{fig:SFRSpectra}(a) shows SFRS spectra for the $\sigma^+/\sigma^+$ and $\sigma^-/\sigma^-$ excitation/detection polarizations in a magnetic field of 2~T.
Two pairs of lines corresponding to transitions where the angular momentum changes by $n=1,2$ are detected. Their magnetic field dependence of the Raman shift
follows the expected linear dependence, $n g \mu_B B$, with $g=2.0$ [see inset in Fig.~\ref{fig:SFRSpectra}(a)], corresponding therefore to spin-flip transitions
between the Zeeman sublevels of the Mn$^{2+}$ ions, i.e. the ionized manganese acceptors $\mathrm{A^{-}}$ \cite{Footnote}. The magnetic field dependence of the SAS
asymmetry parameter $\eta_n^\sigma$ for $\sigma^+$ and $\sigma^-$ excitation and $n=1,2$ is shown in Fig.~\ref{fig:SFRSpectra}(b). The $\eta_n^\sigma$ increase with
$n$ and $B$, and they also depend on the helicity of laser excitation, from which we can conclude that optical orientation of manganese has been accomplished. Using
Eq.~(\ref{eq:SFRSasymmetry}) we deduce the dependence of the manganese polarization $P_M$ on magnetic field for $\sigma^+$ and $\sigma^-$ excitation. The difference
between the two polarizations allows us to estimate the contribution of the optical orientation to the manganese polarization
$P_{OO}=[P_M(\sigma^+,B)-P_M(\sigma^-,B)]/2 \approx -0.25$, independent of magnetic field for $B > 1$~T [see Fig.~\ref{fig:SFRSpectra}(c)]. The other contribution
to $P_M$, related to the thermal population of the Zeeman sublevels $P_B=[P_M(\sigma^+,B)+P_M(\sigma^-,B)]/2$, grows with field and is comparable with $P_{OO}$ at
$B \approx 1$~T. The spin-flip resonance profile has a maximum at about 1.513~eV, which corresponds to resonant excitation of the exciton bound to the charged donor
$\mathrm{D^+X}$ \cite{Footnote3}. Moreover, the most intensive signals appear for pump density around 50~W/cm$^2$. The decrease of the spin-flip scattering
intensity at higher pump densities is related with neutralization of ionized acceptors after capture of photoexcited holes \cite{Astakhov08}.

The resulting Mn$^{2+}$ polarization in longitudinal magnetic field is given by a dynamical equation analogous to the Overhauser effect \cite{Abragam}
\begin{eqnarray}
\label{eq:rateMnPola}    \frac{dP_M}{dt}=-\frac{1}{T_M} \left[ P_M-P_T-\frac{I+1}{S+1}(P_e-P_{eT}) \right] \nonumber \\ -\frac{1}{T_L}(P_M-P_T).
\end{eqnarray}
The first term on the right side describes the dynamical polarization of manganese with time $T_M$ via electrons with polarization $P_e$. The second term reflects
the spin relaxation to the equilibrium value $P_T=-(I+1)\mu_BgB/3k_BT$ with time $T_L$. Here $T$ is the temperature and $k_B$ is the Boltzmann constant. The
equilibrium electron polarization $P_{eT} = -(S+1)\mu_Bg_eB/3k_BT$ is much smaller than $P_T$ since spin $S=1/2$ and Lande factor $g_e=-0.42$ of the electron are
significantly smaller than those for Mn$^{2+}$. The steady state solution of Eq.~(\ref{eq:rateMnPola}) neglecting $P_{eT}$ gives
\begin{equation}
\label{eq:MnPola}    P_M = P_T + \frac{T_L}{T_L+T_M}\frac{I+1}{S+1}P_e.
\end{equation}
Here the first term corresponds to the equilibrium orientation
resulting from thermalization on the Zeeman sublevels (i.e.
$P_T=P_B$), while the second term is the dynamical polarization
$P_{OO}$ by electrons. The electron polarization is determined by
the laser helicity. A polarization $|P_e|=0.36$ was directly
measured by optical orientation at the band edge under the same
experimental conditions. Therefore, Eq.~(\ref{eq:MnPola}) can be
used for evaluating the time ratio $T_M/T_L$ and the actual
temperature $T$ in the region of illumination. Using the
experimental value $P_{OO}= - 0.25$  we obtain $T_M/T_L  = 2$ and
$\mu_BgB/T=0.35$ (for $B=1$~T). This corresponds to $T=4$~K, i.e.
the real temperature under illumination is increased by 2 K. The
same value for $T$ follows from the magnetic field dependence of
$P_B$ when fitted with a Brillouin function, see
Fig.~\ref{fig:SFRSpectra}(c).

Electron spin-flip scattering with Mn$^{2+}$ ions should also manifest itself in the dynamics of the average electron spin. Indeed, since the Mn polarization will
be transferred back into the electron spin-system, the average electron spin should contain information about the Mn polarization. In absence of other relaxation
channels the dynamics of electron polarization is given by an expression similar to the $T_M$-term in Eq.~(\ref{eq:rateMnPola})
\begin{equation}
\label{eq:rateElectrPola}    \frac{dP_e}{dt} = -\frac{1}{\tau_S} \left[ P_e - P_{eT} -\frac{S+1}{I+1}(P_M-P_T) \right],
\end{equation}
since it reflects the conservation of total spin in the flip-flop
process. The electron spin relaxation time $\tau_S$ depends on the
Mn$^{2+}$ concentration $N_{M}$ and time $T_M$ through
$\tau_S=\frac{S(S+1)}{I(I+1)}\frac{N_e}{N_M}T_M$. For small
photoelectron concentrations $N_e \ll N_M$ the electron spin
relaxation time is much shorter than that of manganese, i.e. $\tau_S
\ll T_M$ (a similar relation holds for the electron-nuclear system).
From TRPL $\tau_S \sim 10$~ns and, therefore, one can estimate $T_M
\sim 10~\mu$s for $N_e=10^{14}$~cm$^{-3}$, and $N_M \sim N_D \sim
10^{16}$~cm$^{-3}$. Under pulsed photoexcitation with repetition
period $t_i$ (which satisfies the condition $T_M \gg t_i \gg
\tau_S$), the manganese spin polarization is gradually accumulated
and shows no change within time $t_i$. Therefore, in weak magnetic
fields ($|P_T | \ll 1$) the solution of
Eq.~(\ref{eq:rateElectrPola}) in a time domain $t_i$ can be written
as
\begin{equation}
\label{eq:EPola}  P_e(t) = \frac{I+1}{S+1}P_M + \left[ P_i - \frac{I+1}{S+1}P_M  \right] \exp\left(-\frac{t}{\tau_S}\right).
\end{equation}
At the moment of pulsed excitation $t=0$ the maximum possible electron polarization $P_i=-0.5$ is generated \cite{OO}. Subsequently it decays with a characteristic
time $\tau_S$  until it reaches the plateau, which corresponds to the steady state non-equilibrium manganese polarization. The plateau results from the spin back
flow from manganese to the electrons, providing a long-lived electron spin memory.

\begin{figure}
 \begin{minipage}{8 cm}  \epsfxsize=8.2 cm
  \centerline{\epsffile{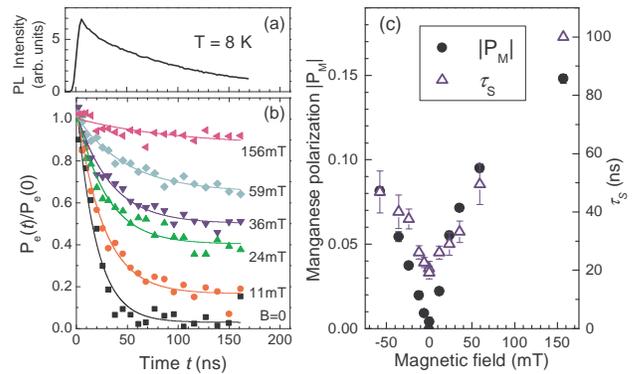}}
  \caption{\label{fig:TPRLplateau}
(a) Intensity decay of the $\mathrm{D^0-A^0}$ PL line. (b) Time evolution of normalized electron polarization for different longitudinal magnetic fields. Solid
lines are fits by Eq.~(\ref{eq:EPola}). (c) Magnetic field dependencies of manganese polarization $P_M$ and electron spin relaxation time $\tau_S$, evaluated from
$P_e(t)$ transients using Eq.~(\ref{eq:EPola}). Pulse repetition period $t_i=1.2~\mu$s.}
  \end{minipage}
\end{figure}

In previous TRPL measurements in weak longitudinal magnetic fields we already observed a slow spin relaxation dynamics (up to $1~\mu$s) of electrons localized on
shallow donors in GaAs:Mn \cite{Akimov10}. However, the small signal-to-noise ratio in these studies did not allow us to draw an unambiguous conclusion on the
non-exponential spin evolution. Large power densities induce undesired local heating. Therefore, we increased the setup sensitivity by enlarging the illumination
area by a factor of 400, while keeping the pulse energy density low at $\mathcal{P}_{exc} \approx 10$~nJ/cm$^2$. The intensity transient of the donor-acceptor
($\mathrm{D^0-A^0}$) PL line is shown in Fig.~\ref{fig:TPRLplateau}(a), showing a decay with 100~ns lifetime.

The decay of circular polarization degree $\rho_c(B,t)= -P_e(B,t)/2$ gives direct access to the spin dynamics of the oriented electrons.
Figure~\ref{fig:TPRLplateau}(b) shows a nontrivial electron spin dynamics \cite{Footnote4}. We find an initial electron spin $|P_e(0)|=0.4$, which is close to the
maximum value of 0.5. After decay within several tens of ns the electron spin polarization reaches a plateau, whose level increases with magnetic field and reaches
0.35 for $B=156$~mT. Using Eq.~(\ref{eq:EPola}) we determine the magnetic field dependencies of $\tau_S$ and $P_M$, as presented in Fig.~\ref{fig:TPRLplateau}(c).
The electron spin relaxation time increases from 20 to 100~ns in a magnetic field of 150~mT. The plateau values corresponding to  $P_M$ are symmetric with respect
to magnetic field inversion and they change sign if the excitation helicity sign is reversed (i.e., they follow the electron spin polarization). This corroborates
our conclusion about optical orientation of manganese. The manganese orientation is absent in zero magnetic field, in accord with \cite{Akimov10}, however, it
appears in weak magnetic fields and saturates for $B>150$~mT in line with the $P_{OO}$ behavior from SFRS, see Fig.~\ref{fig:SFRSpectra}(c).

We can exclude any influence of dynamical polarization of lattice
nuclei since nuclear polarization should appear in much smaller
magnetic fields $B \geq B_L \approx 0.3$~mT, where $B_L$ is the
local field onto a nucleus by the neighboring nuclei \cite{OO}.
Moreover, we do not observe any plateau in p-GaAs samples doped with
non-magnetic Ge acceptors \cite{Akimov10}. Thus, the plateau
evidences manganese dynamic polarization in weak longitudinal
magnetic fields, which suppress Mn$^{2+}$ spin relaxation.

Above we discussed the spin transfer between the localized electrons and the Mn$^{2+}$ ions. This transfer occurs due to fluctuations of the exchange interaction
given by the Hamiltonian $\hat{H}_{sd}^e = -b \mathbf{\hat{S}} \cdot \mathbf{\hat{I}}$. Here $\mathbf{\hat{S}}$ and $\mathbf{\hat{I}}$ are the electron and
manganese spin operators, respectively, and $b$ is a constant depending on the electron probability at the Mn$^{2+}$ site. Apart from the fluctuation term the
exchange interaction in mean-field approximation contains an expression like $-\langle b \rangle I \mathbf{P}_M \cdot \mathbf{\hat{S}} - \langle b \rangle f S
\mathbf{P}_e \cdot \mathbf{\hat{I}}$, where the exchange constant $\langle b \rangle$ is the average electron probability at the Mn$^{2+}$ site and $f$ is the donor
filling factor. The first term describes the interaction of the electron spin with the effective exchange field of manganese $\mathbf{B}_M = - \langle b \rangle I
\mathbf{P}_M/\mu_B g_e$ (analogue of the Overhauser field). The second term corresponds to interaction of manganese spins with the effective field of the electrons
$\mathbf{B}_e = -f \langle b \rangle S \mathbf{P}_e/\mu_B g$ (analogue of the Knight field). For small excitation densities $f \ll 1$, and as a result $B_M \gg
B_e$.

\begin{figure}
 \begin{minipage}{8.2cm}  \epsfxsize=6 cm
  \centerline{\epsffile{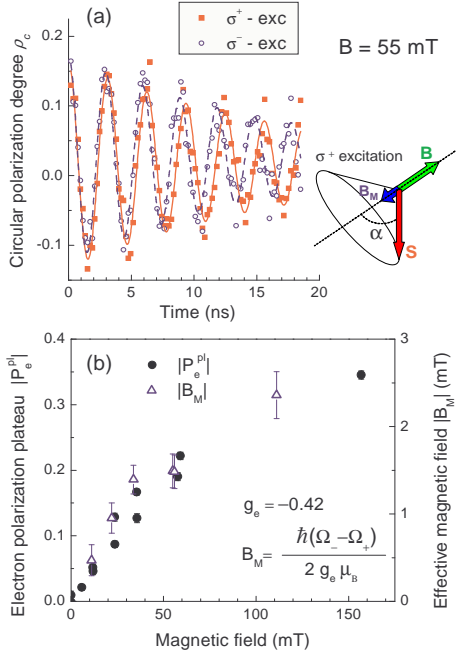}}
  \caption{\label{fig:obliqueB}
(a) Circular polarization oscillations in oblique magnetic field
($B=55$~mT) for excitation with $\sigma^+$ and $\sigma^-$ polarized
photons. $\alpha = 70^\circ$. Inset schematically shows the electron
spin precession in external and exchange fields. (b) Magnetic field
dependence of electron polarization plateau $P_e^{pl}$ and effective
exchange magnetic field $B_{M}$.}
  \end{minipage}
\end{figure}

The presence of the exchange manganese field can be detected via
changes in the electron Larmor precession frequency in an external
magnetic field. Since manganese polarization is absent in Voigt
geometry \cite{Akimov10} it is necessary to perform this experiment
in oblique magnetic field for $\sigma^+$ and $\sigma^-$ polarized
excitation. These data are summarized in Fig.~\ref{fig:obliqueB}.
Figure~\ref{fig:obliqueB}(a) shows the oscillations of the circular
polarization degree of the PL in time for the two opposite
helicities ($B=55$~mT applied at an angle $\alpha \approx 70^\circ$
with respect to the light propagation direction). Indeed we see that
the Larmor frequencies $\Omega_\pm$ corresponding to electron spin
precession with $g_e=-0.42$  are different ($\Omega_- - \Omega_+ =
0.11$~ns$^{-1}$). Based on this difference we can plot the
dependence of the effective magnetic field $B_{M} = \hbar (\Omega_-
-\Omega_+)/2\mu_Bg_e$ on external magnetic field $B$, as shown in
Fig.~\ref{fig:obliqueB}(b). The clear correlation between the
magnetic field dependencies of $B_{M}$ and the plateau value
$P_e^{pl}$ indicates their common origin: both are proportional to
the non-equilibrium manganese polarization $P_M$.

One can also estimate the exchange constant $\langle b \rangle$. We note that in oblique magnetic fields no oscillations with the Mn$^{2+}$ $g$ factor are observed.
This indicates that the manganese orientation is parallel to the direction of the external magnetic field $\mathbf{B}$, i.e. effective optical pumping occurs only
for the spin component along $\mathbf{B}$. Therefore, the polarization $P_M(\alpha)=P_M(\alpha=0)\cos(\alpha)$. Then in the oblique geometry the field $B_M = -
\langle b \rangle I P_M(\alpha=0)\cos(\alpha)/\mu_B g_e$. Simultaneously from Eq.~(\ref{eq:EPola}) we have for the plateau $P_e^{pl} =
\frac{I+1}{S+1}P_M(\alpha=0)$. The ratio of $B_M/P_e^{pl}$ from Fig.~\ref{fig:obliqueB}(b) allows one to derive $\langle b \rangle =0.5~\mu$eV. This is in agreement
with the value of $b_0 =2.4~\mu$eV for Mn$^{2+}$ in the center of the donor, giving an upper limit for $\langle b \rangle$  \cite{Sapega01}.

In conclusion, we have demonstrated optical orientation of Mn$^{2+}$ acceptors in GaAs using two optical techniques. Dynamic manganese polarization is established
in weak longitudinal magnetic fields ($B \leq 100$~mT), which are required to suppress the Mn$^{2+}$ spin relaxation. The optically oriented Mn$^{2+}$ ions maintain
the spin and return part of the polarization back to the electron spin system providing a long-lived electron spin memory.

The authors are grateful to K.~V.~Kavokin for useful discussions.
This work was supported by the Deutsche Forschungsgemeinschaft
(Grant No. 436RUS$113/958/$0-1) and the Russian Foundation for Basic
Research.

\appendix
\section{Derivation of Equation~(\ref{eq:SFRSasymmetry})}

\begin{figure}
 \begin{minipage}{8.2cm}  \epsfxsize=8 cm
  \centerline{\epsffile{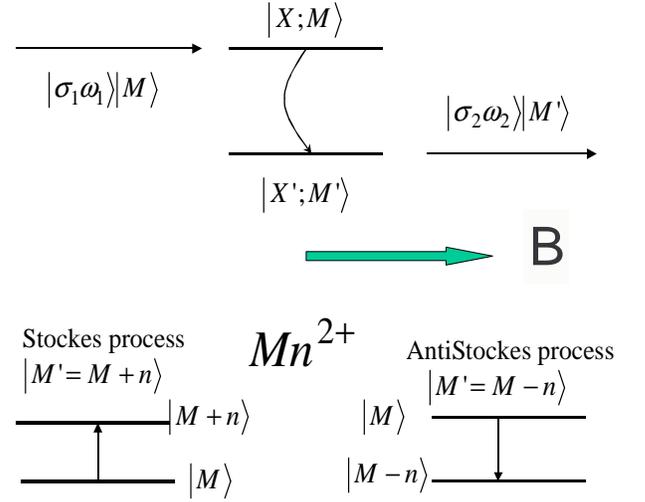}}
  \caption{\label{fig:schema}
Raman scattering of circular polarized light accompanied by spin flip of Mn$^{2+}$ in longitudinal magnetic field.}
  \end{minipage}
\end{figure}

Raman scattering of circular polarized light accompanied by spin flip of Mn$^{2+}$ in longitudinal magnetic field (Faraday geometry) is shown in
Fig.~\ref{fig:schema}. In the initial state $|\sigma_1\omega_1\rangle$  the photon with circular polarization $|\sigma_1\rangle$  and energy quant  $\hbar\omega_1$
propagates along the magnetic field $\mathbf{B}$. There is Mn$^{2+}$ ion in a quantum state $|M\rangle$ with a given spin projection $M=-5/2~\ldots+5/2$ on
$\mathbf{B}$ direction and Zeeman energy $E_M=+\mu_BgBM$ (Bohr magneton $\mu_B>0$, $g$-factor $g=2.0$). In the intermediate state there is an exciton $|X\rangle$,
which due to exchange interaction transfers the manganese from state $|M\rangle$ into $|M^\prime=M\pm n\rangle$ , while changing its state into $|X^\prime\rangle$.
In the final state $|\sigma_2\omega_2\rangle$ the photon with circular polarization $|\sigma_2\rangle$ and energy quant
$\hbar\omega_2=\hbar\omega_1+\mu_BgB(M-M^\prime)$ propagating along the direction $\mathbf{B}$ is registered. In case if the manganese projection is increased
(decreased) by $n$ the photon energy decreases (increases) by $n\cdot\mu_BgB$ with respect to initial energy value, leading to the frequency shift of the scattered
light in Stokes (anti-Stokes) region.

The number of light scattering events (in time per unit volume) in the direction parallel to the magnetic field from the initial into the final state is given by
the probability of scattering $W_{\sigma_1\omega_1M\rightarrow\sigma_2\omega_2M^\prime}$ from $|\sigma_1\omega_1M\rangle$ into $|\sigma_2\omega_2M^\prime\rangle$
and the initial manganese states $|M\rangle$ distribution function $N_M^{\sigma_1}(B)$ in magnetic field under excitation with $|\sigma_1\rangle$ polarized light.
\begin{equation}
\label{eq:A1}    \mathcal{I} = W_{\sigma_1\omega_1M\rightarrow\sigma_2\omega_2M^\prime} N_M^{\sigma_1}(B). \tag{A1}
\end{equation}
The dependence of $N_M^{\sigma_1}(B)$ from excitation light helicity $|\sigma_1\rangle$ is especially important. It takes into account possible optical orientation
of manganese ions. Since the spin-flip is detected at the energy  $\mu_BgB \cdot |M^\prime-M|$ in the final state of the crystal there is only manganese spin, which
is flipped, while all other lattice variables are fixed. Therefore the probabilities for back and forward transitions are equal to each other
\begin{equation}
\label{eq:A2}   W_{\sigma_1\omega_1M\rightarrow\sigma_2\omega_2M^\prime} =   W_{\sigma_2\omega_2M^\prime\rightarrow\sigma_1\omega_1M}. \tag{A2}
\end{equation}
If the spectral width of spin-flip resonance profile is larger than the Zeeman splitting of manganese spin sublevels, then the transition probability weakly depends
on the light frequency $\omega_1 (\omega_2)$. Such kind of situation is realized in GaAs:Mn with $N_{Mn}\sim10^{17}$~cm$^{-3}$. Therefore we assume that
$\omega_1=\omega_2=\omega$ in Eq.~(\ref{eq:A2}) and leave index $\omega$.

Let us show that for identical polarizer and analyzer positions $(\sigma_1=\sigma_2\equiv\sigma)$ the asymmetry parameter $\eta^\sigma_n$ does not depend on exact
form of probabilities $W_{\sigma M \rightarrow \sigma M^\prime }\equiv W_{M,M^\prime} = W_{M^\prime,M}$ for small manganese polarization
$P_M(\sigma,B)=\frac{1}{I}\sum^{+5/2}_{M=-5/2}M\cdot N^\sigma_M(B)$. In this case the detected signal corresponds to the emission of excitons, which do not change
their spin polarization in the scattering process. We consider the spin-flip process of n-th order when $M^\prime = M \pm n$, where $1 \leq n \leq 2I = 5$. Then the
intensity of transitions in Stokes range ($M^\prime = M + n$, photon energy shift to lower energies by $n \cdot \mu_BgB$) for polarizer and analyzer with
polarization $\sigma$
\begin{equation}
\label{eq:A3} \mathcal{I}_{n\sigma}^S = \sum^{+5/2}_{M=-5/2} W_{M,M+n}N_M^\sigma(B). \tag{A3}
\end{equation}
In the same way the intensity of transitions in anti-Stokes range ($M^\prime = M - n$, photon energy shift to higher energies by $n \cdot \mu_BgB$)
\begin{equation}
\label{eq:A4} \mathcal{I}_{n\sigma}^{AS} = \sum^{+5/2}_{M=-5/2} W_{M,M-n}N_M^\sigma(B). \tag{A4}
\end{equation}
We assume that the manganese spin sublevels distribution function
\begin{equation}
\label{eq:A5} N_M^\sigma(B)=\frac{1}{2I+1}\left[ 1+\frac{3M}{I+1}P_M(\sigma,B)\right] \tag{A5}
\end{equation}
is defined only by polarization. It may take place when the manganese spin system is mainly disordered, i.e. $P_M\ll1$ , while the alignment parameters of higher
order are negligible. The equation (1) follows from Eqs. (\ref{eq:A2}-\ref{eq:A5}) and the following relations
\begin{eqnarray}
\sum^{+\frac{5}{2}}_{M=-\frac{5}{2}}W_{M,M-n}=\sum^{+\frac{5}{2}}_{M=-\frac{5}{2}} W_{M-n,M}=\nonumber \\
=\sum^{+\frac{5}{2}-n}_{M^\prime=-\frac{5}{2}-n} W_{M^\prime,M^\prime+n}= \nonumber\\
=\sum^{+\frac{5}{2}-n}_{M=-\frac{5}{2}} W_{M,M+n}=\sum^{+\frac{5}{2}}_{M=-\frac{5}{2}} W_{M,M+n} \nonumber
\end{eqnarray}
and
\begin{eqnarray}
\sum^{+\frac{5}{2}}_{M=-\frac{5}{2}}M \cdot W_{M,M-n} = \sum^{+\frac{5}{2}}_{M=-\frac{5}{2}} M \cdot W_{M-n,M}=\nonumber \\
=\sum^{+\frac{5}{2}-n}_{M^\prime=-\frac{5}{2}-n} (M^\prime + n) \cdot W_{M^\prime,M^\prime+n}= \nonumber\\
=\sum^{+\frac{5}{2}}_{M=-\frac{5}{2}} (M + n) \cdot W_{M,M+n}. \nonumber
\end{eqnarray}
Here we take into account that $W_{M,M^\prime} =0 $  if one of the indexes $M(M^\prime)$  is out of the $[-5/2,+5/2]$  range and  $W_{M,M+n}=W_{M+n,M},
W_{M,M-n}=W_{M-n,M}$.

It follows that the Eq.~(1) does not contain probability $W_{M,M^\prime}$. It is also seen that the asymmetry parameter is proportional to the scattering order $n$.
The Eq.~(1) is deduced for the identical positions of polarizer and analyzer. In case if they are crossed the probabilities of the transitions become important and
the difference in intensities of Stokes and anti-Stokes components may take place even without manganese polarization. For example, $\sigma^+$-photon creates the
exciton with momentum projection $m=+1$ and after interaction with manganese it transforms into the state with $m=-1$ and subsequently emits a $\sigma^-$-photon.
Simultaneously the manganese momentum projection increases by 2 so that photon frequency is shifted into the Stokes region. The anti-Stokes component for such
scattering process does not exist at all. However this is related to selection rules and not to manganese polarization. In order to find the manganese polarization
in case of crossed polarizer and analyzer it is necessary to compare the intensity of Stokes component in $\sigma/\bar{\sigma}$ configuration with the intensity of
anti-Stokes component in the opposite $\bar{\sigma}/\sigma$ configuration, which is not always convenient from experimental point of view.

We emphasize that for  $n>1$ the Eq.~(1) is valid only for a spin-flip of \textit{single} manganese. It is not valid if there are two ions in the exciton
localization region and each of these ions flips its spin resulting in the total projection change of $n$. Therefore the sample should be doped in a such way that
the average number of Mn$^{2+}$ ($I=5/2$) ions in the exciton localization volume is below 1. This corresponds to the case of GaAs doped with Mn. The main part of
manganese are bound with the holes forming the neutral magnetic acceptor with the total spin $F=1$. Due to the partial compensation of the acceptors with shallow
donors ($N_D\sim10^{16}$~cm$^{-3}$) the number of ionized acceptors (without the hole) is comparable with $N_D$. In this case the number of ions in the region of
exciton localization (Bohr radius $a_X = 12$~nm) is not larger than 1 because the parameter $\frac{4\pi}{3}N_Da_X^3\leq0.1$. Note that the mentioned condition can
be violated for the spin flip of magnetic acceptors ($F=1$), which have significantly larger concentration of $N_{A^0}=8\cdot10^{17}$~cm$^{-3}$  in the studied
samples. In exciton localization volume there there are 2-3 of such acceptors, which allows the combined spin-flip processes.


\begin{references}
\bibitem{Footnote}
No SFRS signal is observed for neutral manganese acceptors $\mathrm{A^0}$  ($g_A$=2.74).
This may be due to local stress, electric fields or fast spin
relaxation of the $\mathrm{A^0}$ complex, smearing out the resonance
\cite{Sapega01}.
\bibitem{Footnote3}
Excitation of the energetically close lying $\mathrm{A^0X}$ complex
does not influence the SFRS signal on the ionized manganese
$\mathrm{A^-}$ since the exciton bound to a neutral acceptor
$\mathrm{A^0}$ is localized far enough from a neighboring
$\mathrm{A^-}$ ($\frac{4\pi}{3}N_M a_B^3 \ll 1$, $a_B$ is the donor
Bohr radius). Simultaneously all $\mathrm{A^-}$ energetically prefer
to stay near the ionized donors $\mathrm{D^+}$. Therefore SFRS on
$\mathrm{A^-}$ must be related with $\mathrm{D^+X}$ excitation. Note
that the $\mathrm{A^-X}$ complex is not stable in bulk GaAs.
\bibitem{Footnote4}
SFRS is related to excitation of a $\mathrm{D^+X}$ located closely to an ionized acceptor $\mathrm{A^-}$. TRPL shows a long lifetime of the $\mathrm{D^0-A^0}$
transition, during which the electron undergoes several hundreds of jumps between the donors. On its way the electron can approach an ionized acceptor and stimulate
the spin-flip transition. Therefore, spin polarization of $\mathrm{A^-}$ ions will occur for both SFRS and TRPL. Note, that similar flip-flop process will take
place at the neutral acceptors, whose number is approximately 60 times larger, although the exchange interaction with them is smaller due to screening effects
originating from the anti-ferromagnetic interaction between a hole and a manganese spin \cite{Astakhov08}. However SFRS signal related to $\mathrm{A^0}$ is not
observed in the studied sample \cite{Footnote}.
\end{references}
\end{document}